\documentstyle[sprocl,epsfig]{article}

\def\NPA{{\em Nucl. Phys.} A}

\def\PLB{{\em Phys. Lett.}  B}
\def\PRL{{\em Phys. Rev. Lett.}}
\def\PRD{{\em Phys. Rev.} D}

\def\be{\begin{equation}}
\def\ee{\end{equation}}
\def\bea{\begin{eqnarray}}
\def\eea{\end{eqnarray}}

\begin{document}

\title{MULTIPARTICLE PRODUCTION AND PERCOLATION OF STRINGS\footnote{Talk given
in the workshop {\it Particle Distributions in Hadronic and Nuclear
Collisions}, University of Illinois at Chicago.}}

\author{\underline{C. Pajares}, C.A. Salgado}

\address{Departamento de F\'\i sica de Part\'\i culas\\ Universidade de
Santiago de Compostela\\ E-15706 Santiago de Compostela, Spain}

\author{E.G. Ferreiro}

\address{Laboratoire de Physique Th\'eorique et Hautes Energies\\ Universit\'e
de Paris-Sud, B\^atiment 211, F-91405 Orsay Cedex, France.}


\maketitle\abstracts{It is shown that the multiplicity distribution associated
to self-shadowing events satisfies an universal equation in terms of the minimum
bias distribution. The number of elementary collisions, basic in the structure
of any multiple scattering, is reduced if a collective effect like percolation
of strings takes place. The main consequences of the percolation are briefly
discussed.} 

\section{Introduction}
Most of the exciting and forthcoming data on particle production on
hadron-hadron and nucleus-nucleus collisions can be understood by conventional
physics. This does not mean lack of excitement. Indeed, in this paper we show
first as an example a surprising general law on multiplicity distributions
associated to events which are shadowed by themselves, obtained using nothing
but unitarity \cite{1}.

Also, some data of the forthcoming experiments in the Relativistic Heavy Ion
Collider (RHIC) and in the Large Hadron Collider (LHC) can give us many
surprises related to new physics. As an example of this, some
effects of the percolation of strings \cite{2,3}, that can be detected already
at RHIC, are discussed in this paper.

\section{Multiplicity Distributions Associated to Events Shadowed by
Themselves.}
A multiple collision can be seen as a superposition of many elementary
collisions, and it is worth to ask for the propagation of properties existing
at the level of elementary collisions. In this way, for example, it is well
known the $A-$behavior of the hadron-nucleus cross sections for events which
are shadowed by themselves \cite{4,5}. Indeed, for these events, denoted by
C-events, the resulting $h-A$ event is of type C only and only if at least one
of the elementary collisions if of type C. It is said, the algebra

\bea
C\times \mbox{no }C&=&C\nonumber\\
\mbox{no }C\times \mbox{no }C&=&\mbox{no }C\nonumber\\
C\times C&=&C\nonumber
\eea

\noindent
is satisfied

From the inelastic cross section
\be
\sigma_{in}^{hA}=1-(1-\sigma T(b))^A=\sum_{n=1}^A(\sigma T(b))^n(1-\sigma
T(b))^{A-n}
\label{1}
\ee

\noindent
retaining in the sum
\be
(\sigma T(b))^n=\sum_{i=0}^n
\Big(\begin{array}{c} n\\i\end{array}\Big)
\sigma_C^i\sigma_{\mbox
{no}C}^{n-i}T(b)^n 
\label{2}
\ee

\noindent
the terms with $i\geq 1$, the C-cross section is obtained:
\be
\sigma^{hA}_C=1-(1-\sigma_CT(b))^A
\label{3}
\ee

If $\alpha_C$ is small, $\sigma^{hA}_C$ behaves as $A\sigma_C$, otherwise
it behaves as $A^{2/3}\sigma_C$.

Let us next consider the particle distribution associated to a rare event 
C (Drell-Yan pairs, annihilation cross section,
particles of a given type produced in a rapidity interval, slow nucleons, high
$p_T$ particle, etc.)

If $\alpha_C$ is the probability of event C to occur in an elementary collision
and $N(\nu)$ is the number of events with $\nu$ elementary collisions in a
nucleus-nucleus collision, the number of events where event C occurs is given
by

\be
N_C(\nu)=\sum_{i=1}^\nu
\Big(\begin{array}{c} \nu\\i\end{array}\Big)
\alpha_C^i(1-\alpha_C)^{\nu-i}N(\nu)
\label{4}
\ee

If event C is rare ($\alpha_C$ small), we can retain the leading term in
$\alpha_C$ and

\be
N_C(\nu)\simeq \alpha_C\nu N(\nu)
\label{5}
\ee

The total number of C-events and total number of events are:

\bea
\sum N_C(\nu)&=&\alpha_C\langle \nu\rangle N\nonumber\\
\sum N_(\nu)&=&N\label{6}
\eea
\noindent

Therefore
\be
P_C(\nu)={\nu\over\langle\nu\rangle}P(\nu)
\label{7}
\ee

The multiplicity distribution on produced particles $P(n)$ is given by

\be
P(n)=\sum_{\nu=1}\sum_{n_1+\dots+n_\nu=n} \varphi(\nu)p(n_1)p(n_2)\dots
p(n_\nu)
\label{8}
\ee
and the different moments of the multiplicity distribution can be expressed in
terms of the different moments of the distribution on the number of elementary
collisions and of the distribution on the particles produced in the elementary
collision. In particular:

\bea
\langle n\rangle &=&\langle \nu\rangle \bar n\nonumber\\
{{D^2}\over{\langle n\rangle ^2}}&=&{{\langle \nu^2\rangle -\langle \nu\rangle
^2}
\over
{\langle \nu\rangle ^2}}+{{1}\over{\langle \nu\rangle }}
{{d^2}\over{\bar n^2}}\label{9}
\eea

${{D^2}\over{\langle n\rangle ^2}}$ increases with the complexity of the
systems involved, from 0.09 in $e^+e^-$ to 0.25-0.3 in $p\bar p$ collisions and
to 0.8-1 in $AB$ collisions. 
In general, the main contribution to the different normalized moments of the
total multiplicity distribution is given by the normalized moments of the
distribution on the number of elementary collisions, in such a way that in
(\ref{7}) $\nu$ can be exchanged by $n$ or $E_T$, the transverse energy
(experimentally $n$ depends linearly on $E_T$):

\bea
P_C(n)={n\over\langle n\rangle }P(n),\ \ \ \ \ 
P_C(E_T)={E_T\over \langle E_T\rangle }P(E_T)\label{10}
\eea

Relations (\ref{10}) are universal, independent of $\alpha_C$. In figure 1 our
prediction for Drell-Yan pairs according to (\ref{10}) is compared with the
NA38, $SU$ data \cite{6}. Similar good agreements are obtained for many other
cases like high $p_T$ particles produced in $\alpha\alpha$ collisions, $W^\pm$
and $Z^0$ production in $p\bar p$ collisions (see figure 2) or $p\bar p$
annihilations, as can be seen in reference 1. 

\begin{figure}[t]
\begin{center}
\epsfig{figure=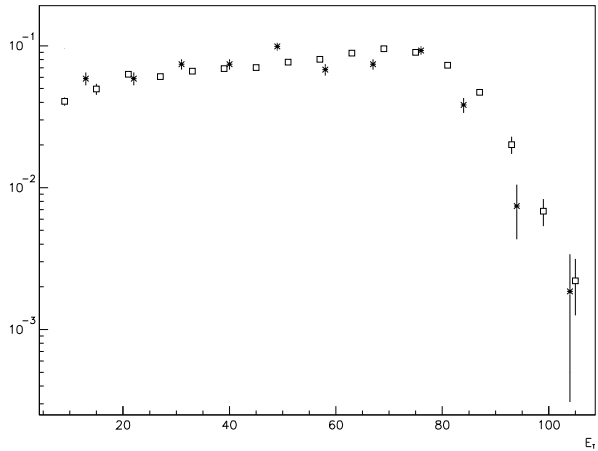,height=7cm}
\caption{NA38 experimental associated multiplicity to Drell-Yan pairs in S-U
collisions (cross points) compared to ${n\over <n>}P(n)$ where $P(n)$
is the experimental multiplicity distribution of S-U
collisions (squared points).}
\end{center}
\end{figure}

\begin{figure}[t]
\begin{center}
\epsfig{figure=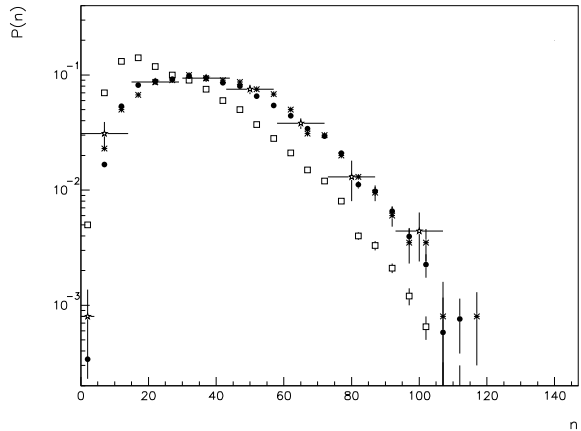,height=7cm}
\caption{Prediction for the associated multiplicity distribution for
$W^{\pm}$ and $Z^0$ events (round black points) together with the experimental
data ($W^{\pm}$ cross points and $Z^0$ white stars) and the minimum bias
distribution (squares).}
\end{center}
\end{figure}

\subsection{$J/\Psi$ Suppression}

In the case of $J/\Psi$ production, final state destructive absorption will
eliminate the linearity dependence of $N_C(\nu)$ on $\nu$ by making the
effective number of collisions where event C appears smaller. This means that
instead of equation (\ref{6}) we have

\be
N_C(\nu)=\alpha_C\nu^\varepsilon N(\nu)\ \ , \varepsilon < 1
\label{11}
\ee

\noindent
and
\be 
P_C(E_T)={E_T^\varepsilon\over\langle E_T^\varepsilon\rangle } P(E_T)
\label{12}
\ee

In figure 3 it is seen that the $E_T$ distribution associated to $J/\Psi$ in
$SU$ collisions can be described by formula (\ref{12}) with $\varepsilon=0.7$.

\begin{figure}[t]
\begin{center}
\epsfig{figure=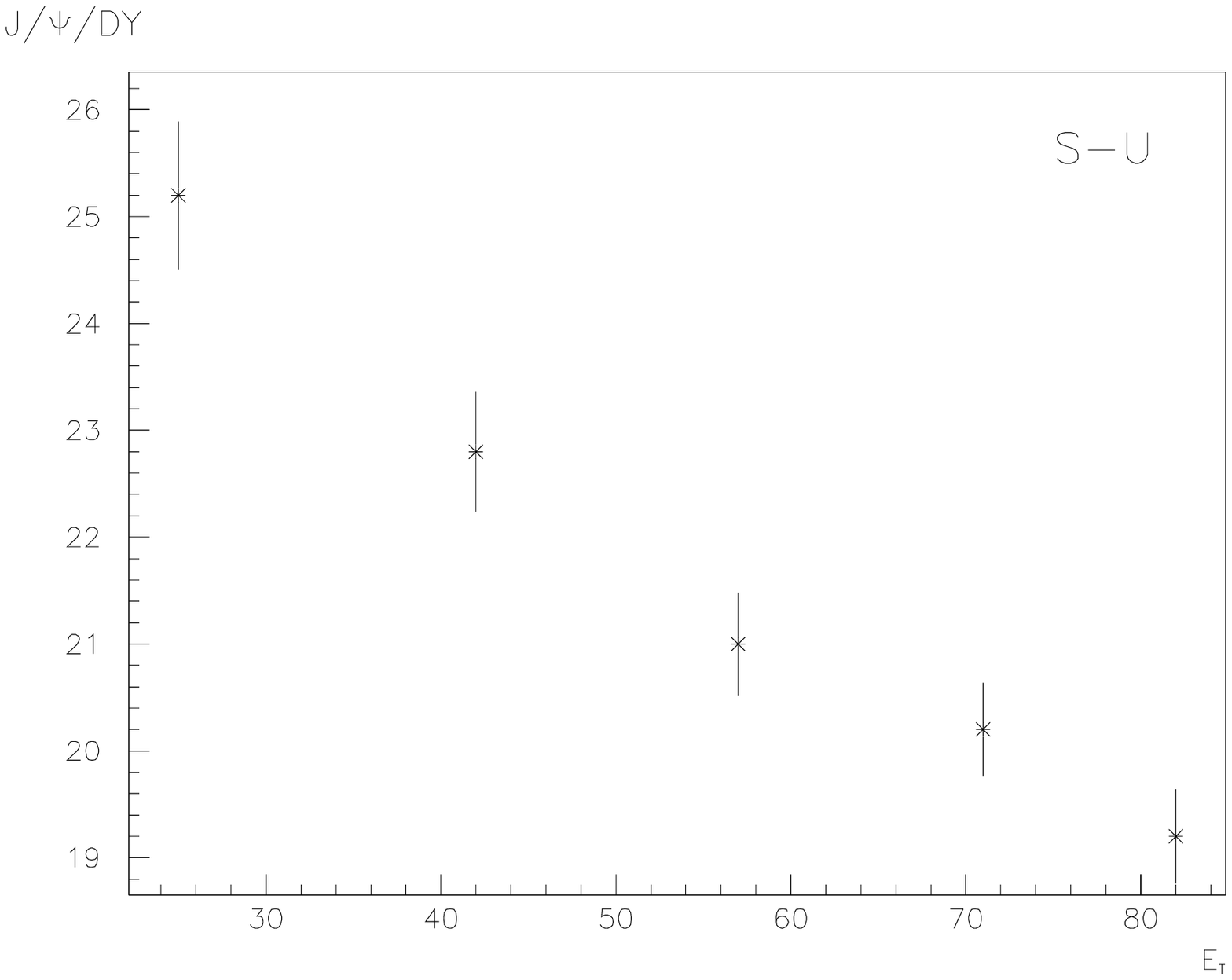,height=7cm}
\caption{Experimental $J/\Psi$ over DY pairs as a function of $E_T$ in S-U
collisions.}
\end{center}
\end{figure}

Notice that from (\ref{10}) and (\ref{11})
\be
N_{J/\Psi}(E_T)/N_{DY}(E_T) \sim 1/E_T^{1-\varepsilon}
\label{13}
\ee

This ratio decreases with $E_T$ and the curvature, the second derivative, is
always positive. In all computations of $J/\Psi$ absorption, including
destruction by comovers, the tendency for a large $E_T$ saturation occurs,
which implies positive curvature.

There is, however, another possibility for changing the $\nu$ linearity of
equation (\ref{6}): in a Quark Gluon Plasma (QGP), the $J/\Psi$ and $\Psi'$ 
formation will be prevented \cite{7}. Now $\alpha_C$ is a function of $\nu$
vanishing for large $\nu$. If the transition is very sharp:

\bea
\alpha_{J/\Psi}(\nu)=\alpha_{J/\Psi}\ \ \, & \nu\leq\nu^*\ \ \, \nonumber \\
\alpha_{J/\Psi}(\nu)=0\ \ \, &  \nu > \nu^*\ \ \, \label{14}
\eea
\noindent
or making a more reasonable approximation

\be
\alpha_{J/\Psi}(\nu)=\alpha_{J/\Psi} \exp(-\nu^2/\nu^{*^2})
\label{15}
\ee
Now
\be
N_{J/\Psi}(E_T)/N_{DY}(E_T) \sim \exp(-E_T^2/E_T^{*^2})
\label{16}
\ee

\noindent
which presents an essential difference to (\ref{13}). For $E_T>E_T^*$ the ratio
has a negative curvature. The experimental ratio for $SU$ does not present any
change in the curvature. At first sight, the NA50 experimental data in
$PbPb$ \cite{8} presents a change in the curvature for central events. However,
it has been shown that it can be described without any change \cite{9} using
formula (\ref{12}) with $\varepsilon=0.6$. This value is obtained previously
from the low $E_T$ associated probability. 

Coming back to our basic relation (\ref{16}) we can mention another possibility
for distinguish absorption or comovers effects from QGP formation. In the first
case $P_{J\Psi}(E_T)\geq P(E_T)$ when $E_T\to \infty$, while if plasma is
formed $P_{J\Psi}(E_T)< P(E_T)$. The $SU$ data are consistent with the first
inequality.

Independently, whether or not, QGP is formed, it is worth to ask for some
reason to distinguish central $PbPb$ collisions at SPS energies from the rest
of collisions. The answer is percolation of strings \cite{2}.

\section{Percolation of Strings}
In many models of hadronic collisions, color strings are exchanged between
projectile and target. The number of strings grows with the energy and with the
number of nucleons of the participating nuclei. When the density of strings
becomes high, the string color fields begin to overlap and eventually
individual string may form a new string which has a higher color charge at its
ends corresponding to the $SU(3)$ sum of the color charges located at the ends
of the original strings. As a result, the higher color means a larger string
tension and heavy flavor is produced more efficiently \cite{10,11}. Also, as
the effective number of strings decreases, the mean multiplicity is less than
in the case of independent string fragmentation.

In impact parameter space these strings are seen as circles inside the total
interaction area. As the number of strings increases, more strings overlap.
Above a critical density of strings percolation occurs, so that paths of
overlapping circles are formed through the whole collision area \cite{2,12}, see
figure 4. Along these paths the medium behaves like a color conductor. The
percolation is a second order phase transition and give rise to a
non-thermalized QGP on a nuclear scale. The percolation threshold $\eta_c$ is
related to the critical density of circles $n_c$ by the expression
\be
\eta_c=\pi r^2 n_c
\label{17}
\ee

\noindent
where $\pi r^2$ is the transverse surface of an string.

\begin{figure}[t]
\begin{center}
\epsfig{figure=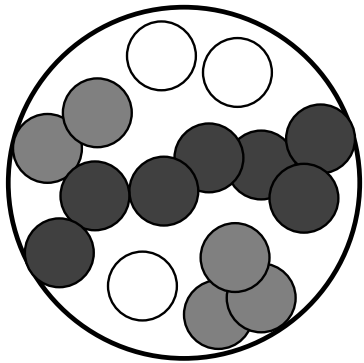,height=5cm}
\caption{Percolation of strings in transverse space.}
\end{center}
\end{figure}

$\eta_c$ has been computed using Monte-Carlo simulation, direct-connectedness
expansions and other methods. All the results are in the range
$\eta_c=1.12-1.18$. Taking $r=0.2\ fm$ which reproduces the $\bar\Lambda$
enhancement observed in $AB$ collisions \cite{2}, it is obtained

\be
n_c=8.9-9.3 \ \mbox{strings}/fm^2
\label{18}
\ee

In table 1 the number of strings exchanged for central $pp$, $SS$, $FeAir$,
$SU$ and $PbPb$ collisions is shown together with their densities. It is seen
that at SPS energies only the density reached in central $PbPb$ collisions is
above the critical density. At RHIC, for central $AgAg$ collisions, the density
of strings is 9, just at the critical value. At LHC energies, $SS$ central
collisions are already over the critical density. Let us mention that for
central $FeAir$ collisions, the critical density is reached at LHC energies. In
the laboratory frame, this means that between $10^{17}$ eV and $10^{18}$ eV the
percolation of strings occurs in central $FeAir$ collisions. Several
experiments of cosmic rays have pointed out a change on the behavior of the
atmospheric cosmic ray cascade around the energies mentioned above. The maximum
depth of the cascade becomes larger indicating that the energy of the primary
is dissipated slower. This is what it was expected if percolation
occurs \cite{13}. The multiplicity of $FeAir$ is suppressed and therefore the
depth of the cascade increases. The usual belief is that a change of the
chemical composition of the primary occurs, and as the energy increases in the
range $10^{16}$ eV to $10^{19}$ eV there are less $Fe$ and more protons in the
primary cosmic ray. Instead of this we pointed out the possibility that the
hadronic phase transition takes place without requiring any ad-hoc change of
the chemical composition of the cosmic ray.

\begin{table}[t]
\caption{Number of strings (upper numbers) and their densities (fm$^{-2}$) (lower
numbers) in central $pp$, $SS$, $SU$ and $PbPb$ collisions at SPS, RHIC and
LHC energies.}
\vspace{0.2cm}
\begin{center}
\footnotesize
\begin{tabular}{|c|c|c|c|c|c|}
\hline
$\sqrt{s}$(AGeV) & \multicolumn{5}{c|}{Collisions}\\
\cline{2-6}
 & $pp$ & $SS$ & $FeAir$ & $SU$ & $PbPb$ \\
\hline
19.4 & 4.2 & 123 & 89 & 268 & 1145 \\
& 1.3 & 3.5 & 4.42 & 7.6 & 9.5 \\
\hline
200 & 7.2 & 215 & 144 & 382 & 1703 \\
& 1.6 & 6.1 & 7.16 & 10.9 & 14.4 \\
\hline
5500 & 13.1 & 380 & 255 & 645 & 3071 \\
& 2.0 & 10.9 & 12.67 & 18.3 & 25.6 \\
\hline
\end{tabular}
\end{center}
\end{table}

Let us point out that if percolation occurs, the energy-momentum of the large
cluster of strings is the sum of the energy-momentum of each individual string,
and therefore when this object fragments it will produce particles with very
large longitudinal momentum, outside of the nucleon-nucleon kinematical limits.
This cummulative and spectacular effect could be checked by Brahms detector at
RHIC. The percolation in this way behaves like a powerful accelerator.

In 2-dimensional percolation it is known that the fraction $\phi$ of the total
area occupied by strings is determined by the formula \cite{14}

\be
\phi=1-\exp(-\eta)
\label{19}
\ee

In the vicinity of the phase transition it is satisfied the scaling law for the
number of clusters with $n$ strings \cite{15}

\be
\langle\nu_n\rangle=n^{-\tau}F(n^\sigma(\eta-\eta_c))\ \ \ \ \ \ 
|\eta-\eta_c|\ll 1\ \ \ \ \ n\gg 1
\label{20}
\ee
\noindent
where the critical indices are $\tau=187/91$ and $\sigma=36/91$. The function
$F(z)$ is finite at $z=0$ and falls off exponentially for $|z|\to\infty$. The
equation (\ref{20}) is of limited value, since near $\eta=\eta_c$ the bulk of
the contribution is still supplied by low values of $n$, for which (\ref{20})
is not valid. However from (\ref{6}) one can finds non-analytical parts of other
quantities of interest at the transition point. In particular, one finds a
singular part of the total number of clusters $M=\sum\nu_n$ as
\be
\Delta\langle M\rangle=c|\eta-\eta_c|^{8/3}
\label{21}
\ee

This singularity is quite weak: not only $\langle M\rangle$ itself but also its
two first derivatives in $\eta$ stay continuous at $\eta=\eta_c$ and only the
third blows up as $|\eta-\eta_c|^{-1/3}$. This singularity implies also a change
on the behavior of the multiplicity. The strength of this singularity would
depend on the dependence \cite{2} of the multiplicity of a cluster on the number
of strings. Reasonable assumptions, imply that the second derivative of the
multiplicity blows up.

\section{Conclusions}

It is well known that the structure of multiple scattering and the shadowing
can explain many unexpected features of experimental data. Here, as an example
of this, it is shown that the self-shadowing events have an universal
associated multiplicity distribution independently of the detailed type of
events, as far as their cross section is small.

Although most of the experimental data of present and may be future heavy ion
collisions can be explained by conventional physics, it is sure that there will
be surprises and excitement related to the forthcoming RHIC and LHC. As an
example, we discuss the possibility of percolation of strings. If this occurs,
exotic phenomena should appear like the production of many very fast particle
with longitudinal momentum much larger that the permitted by the nucleon-nucleon
kinematics or a dumping on the expected multiplicities. It is pointed out, that
the physics of cosmic-ray and the new accelerators begin to overlap. In this
sense, the observed change in the depth of maximum atmospheric shower could be
related to the percolation of strings formed in iron-air collisions.

In conclusion we thank N. Armesto, M. Braun and J. Dias de Deus who
participated in different parts of the work reported here. We would like to
thank the organizers of this workshop for such a nice meeting and the CICYT of
Spain for financial support.

\end{document}